\documentclass[twocolumn,amssymb, nobibnotes, aps, amsmath, amssymb, superscriptaddress,floatfix,]{revtex4-1}

\usepackage[utf8]{inputenc}
\usepackage[T1]{fontenc}
\usepackage{graphicx}
\usepackage{dcolumn}
\usepackage{bm}
\usepackage{newtxtext, newtxmath}  
\DeclareGraphicsExtensions{.pdf,.jpeg,.jpg, .png, .eps, .tiff}
\usepackage{epstopdf}
\usepackage{lipsum}
\usepackage{tabularx}
\usepackage{booktabs}
\usepackage{array}
\usepackage{natbib}
\usepackage[dvipsnames]{xcolor}
\usepackage{grffile}
\usepackage{boxhandler}
\usepackage{float}
\usepackage[normalem]{ulem}
\usepackage{hyperref}
\usepackage{wasysym}
\usepackage{soul}
\usepackage{siunitx}

\renewcommand{\vec}[1]{\boldsymbol{#1}}

\newcommand{\ucmDE}{Departamento de Estructura de la Materia, Física Térmica y Electrónica, Universidad Complutense de Madrid, Madrid, Spain.}
\newcommand{\ucmDQF}{Departamento de Química Física, Facultad de Química, Universidad Complutensede Madrid, Madrid, Spain.}
\newcommand{\GISC}{GISC - Grupo Interdiplinar de Sistemas Complejos, Madrid, Spain.}

\definecolor{pumpkin}{rgb}{1.0,0.4,0.0}
\definecolor{mygreen}{rgb}{0.0,0.55,0.3}
\definecolor{strawberry}{rgb}{1.0,0.0,0.5}
\definecolor{midnight}{rgb}{0.003921569,0.098039216,0.576470588}
\definecolor{saphire}{rgb}{0.0,0.196,0.372549}
\definecolor{crimson}{rgb}{0.75686,0,0.262745}
\definecolor{capri}{rgb}{0.0,0.768627,0.8745098}

\newcommand{\tocite}[1]{{\color{cyan}XXX}}

\newcounter{siequation}
\newcounter{sifigure}
\newcounter{sisection}
\newcounter{simovie}
\newcommand{\ro}[1]{\textcolor{purple}{#1}}

\begin{document}

\title{Phase behavior of a two-dimensional suspension of active pentagons}
\author{R. Fernández-Quevedo García}
\affiliation{\ucmDE}\affiliation{\GISC}
\author{B. Wu-Zhang}
\affiliation{\ucmDE}\affiliation{\ucmDQF}\affiliation{\GISC}
\author{C. M. Barriuso G.}
\affiliation{\ucmDE}\affiliation{\GISC}

\author{W. Lin}
\affiliation{Department of Mechanical Engineering, University of Michigan, Ann Arbor, MI 48109, USA}
\author{M. Ruiz-García}
\affiliation{\ucmDE}
\author{H. Xiao}
\affiliation{Department of Mechanical Engineering, University of Michigan, Ann Arbor, MI 48109, USA}
\author{C. Valeriani}
\affiliation{\ucmDE}\affiliation{\GISC}

\begin{abstract}
Applying  insights acquired  when studying idealized particle systems to real-world amorphous materials remains a significant challenge, primarily due to the complex interplay between particles' shape and interaction mechanisms. It is known that regular pentagons do not tessellate the two dimensional plane, making them the simplest regular polygons unable to form periodic tiling. This geometric constraint, combined with the complexity of 
interactions among non-spherical particles, may give rise to rich emergent behaviors, especially when dealing with self-propelled pentagons. 
Thus, geometry, interactions, and self-propulsion collectively lead to novel dynamical emerging phases. In this context, we investigate how activity and excluded-volume interactions
affect the collective behavior of a two-dimensional suspension of active pentagon-shaped particles. Our results reveal that active pentagons not only undergo  an arrested motility-induced phase separation (MIPS) but, under given conditions, they form  traveling/rotating clusters exhibiting translational/chiral motion.  
Interestingly, both phases present an 
enhanced stability due to particles' activity, shape and excluded-volume interactions, \textcolor{black}{forming an ice-ray lattice whose internal structure represents the densest packing fraction a two-dimensional pentagon suspension may achieve.}

\end{abstract}

\maketitle

\section{Introduction }

Systems made of active units~\cite{Vrugt_2026_colloqium,Vrugt_2025_metareview}, like microorganism suspensions, flocks of birds, bacterial colonies, schools of fish or swarm robots~\cite{Marchetti_2013_review, Bechinger_2016_review}, are common in Nature. These  are all out-of-equilibrium systems, since the active agents are capable of converting an internal energy source into   
forces, leading to persistent motion \cite{elgeti_2015_physics,gompper_2020_2020}. They present emergent collective behaviors which cannot be described by equilibrium statistical physics. A suspension of Active Brownian Particles (ABP) has been shown  to undergo  a Motility Induced Phase Separation (MIPS) or a phase of transient clusters, depending on the inter-particle interactions,  activity, and system density
\cite{Cacciuto_2016, Velasco_2009, jiang2022nonequilibrium, tailleur2008statistical, cates2015motility, redner_2013_structure,levis_2014_clustering, buttinoni_2013_dynamical, redner2016classical, speck2014effective, speck2015dynamical, takatori2014swim,  tjhung2018cluster, zottl2014hydrodynamics, Cates_2015, mandal2019motility, su2021inertia, du2020rod, caprini2020spontaneous, caprini2020hidden}.

In the case of repulsive interactions, 
increasing the activity and density drives the spontaneous aggregation of active particles into locally denser regions, where their persistent motion becomes hindered~\cite{cates2015motility}. 
At the end of this process,  a dense phase of slow, trapped particles coexists with a dilute phase of fast particles: a Motility Induced Phase Separation occurs in the system
\cite{thompson_2011_lattice, fily_2012_athermal, redner_2013_structure, levis_2014_clustering, buttinoni_2013_dynamical}. A complete state diagram for a two dimensional suspension of repulsive active Brownian particles has been reported by 
Digregorio et al \cite{digregorio_2018_full}. Interestingly,  MIPS does not only appear in two dimensional suspensions of spherically active particles, but also in suspensions of particles of different  interactions \cite{caprini2020hidden, wu2024collective, qi2022emergence, du2020rod, zottl2014hydrodynamics,speck2015dynamical, speck2014effective, du2020rod}, and in three dimensional suspensions \cite{wu2024collective, qi2022emergence, theers2018clustering}. 




When considering the  case of attractive interactions, a two dimensional suspension of  ABPs at steady state self-assemble into a  phase of dynamic clusters  \cite{theurkauff2012dynamic, palacci2013living,tanaka_2025}. 
Due to an interplay between the active and the attractive force, a full phase separation cannot take place and activity acts as an effective repulsion between the particles \cite{buttinoni_2013_dynamical, fily_2012_athermal, mognetti2013living}. 

So far, to understand the phase behavior of suspensions of isotropic  active particles, the main focus has been on spherical particles. 
However, more recent works have shifted toward passive and active anisotropic particles such as 
triangles \cite{Martinez_2024, Martinez_2018, Moran_2022, Graaf_2016, moran2022particle}, rectangles \cite{martinez_2006_orientational, Eckert_2025, Moran_2022, avendano_2012_phase, moran2022particle}, pentagons 
\cite{Zhao_2019, henley1986sphere, kuperberg1990double, hales2016packings, moran2022particle}, hexagons and more complex shapes \cite{Moran_2022,hou_2019_phase,hou2024cooperative, moran2022particle,tanaka2025_2}. 

A pentagon is the simplest polygon which cannot tile a two-dimensional planar surface. In a low density suspension, pentagons tend to  form 
local six-fold structures. However, when the density increases, the five-fold symmetry strongly interferes with the hexagonal ordering. Thus, the equilibrated phase has to balance both spatial and orientational degrees of freedom. 
Henley \cite{henley1986sphere} and Kuperberg \cite{kuperberg1990double} proposed several conjectures on 
the highest packing fraction that congruent, regular pentagons could achieve in the Euclidean plane and the optimal packing of regular pentagons. These conjectures were finally proved by Hales \textcolor{black}{and Kusner} \cite{hales2016packings}, suggesting that the optimal packing fraction can be found at approximately $0.92$.

The phase behavior of a two dimensional system of passive pentagon-like particles was first addressed by Branka and coworkers by means of mechanical and computer simulations \cite{branka1991monte,branka1993monte}. 
Rather than pentagons, the authors studied, via Monte-Carlo simulations, 
hard pentamers (particles made of five circular disks arranged with their centers on the vertices of a pentagon)  identifying three phases: a fluid phase, a rectangular crystalline phase, and a hexagonal crystalline phase.
On the other hand, Schilling and Frenkel were the first to unravel the phase behavior of a system of Brownian pentagons \cite{schilling2005monte}. By means of Monte Carlo simulations, they elucidated the effect of the pressure on the positional and orientational order of the pentagons. 
A more recent work \cite{Zhao_2019} studied the jamming transition behavior of  monodisperse photo-elastic pentagons and disks, revealing that jammed pentagons display a local six-fold bond orientational order comparable to their disk counterpart, while sharing similar degrees of translational order. 

One might wonder how activity might affect the phase behavior of repulsive pentagons: whether their phase behavior resembles that of spherical Active Brownian Particles, or their shape impedes close packing~\cite{Berryman_1983,Torquato_2014} thus altering a possible MIPS formation. 
When dealing with active anisotropic particles, Moran and coworkers \cite{moran2022particle} have recently studied a family of translationally self-propelled two-dimensional polygons with a number of edges ranging from three to eight and mixtures of these self-propelled polygons \cite{Moran_2022}. 
Studying only one activity value, the authors concluded that anisotropy, induced by particles' shape, was affecting phase separation and 
the critical density of the MIPS transition. 

Going beyond the work by \cite{moran2022particle},  we develop an active Brownian pentagon model and study 
both structural and dynamical properties of two dimensional systems of active pentagons, while varying the activity (quantified by the Péclet number $\text{Pe}$) and packing fraction ($\rho$). 
 We find that the system forms a homogeneous phase  when  activity  is relatively  low ($\text{Pe} \leq 10$), while clustering (low packing fractions) or MIPS (high packing fractions) phase occurs at higher activities.
 Further structural and dynamical characterizations show that the interplay between the geometry of pentagons and the persistent self-propelling force limits the rotational diffusion of pentagons, stabilizing the resulting aggregates.

 The article is structured as follows. In Section \ref{sec: numerical_details} we include the details of the pentagon model and the analysis tools  used for characterization. In Section \ref{sec: results} we present the results on the phase behavior,  structural properties and  dynamical properties. Finally, we present the conclusions in Section \ref{sec: conclusions}.

\section{Numerical details} \label{sec: numerical_details}

\subsection{A pentagon and its equations of motion}

\begin{figure}[h!]
    \centering
    \includegraphics[width=1\linewidth]{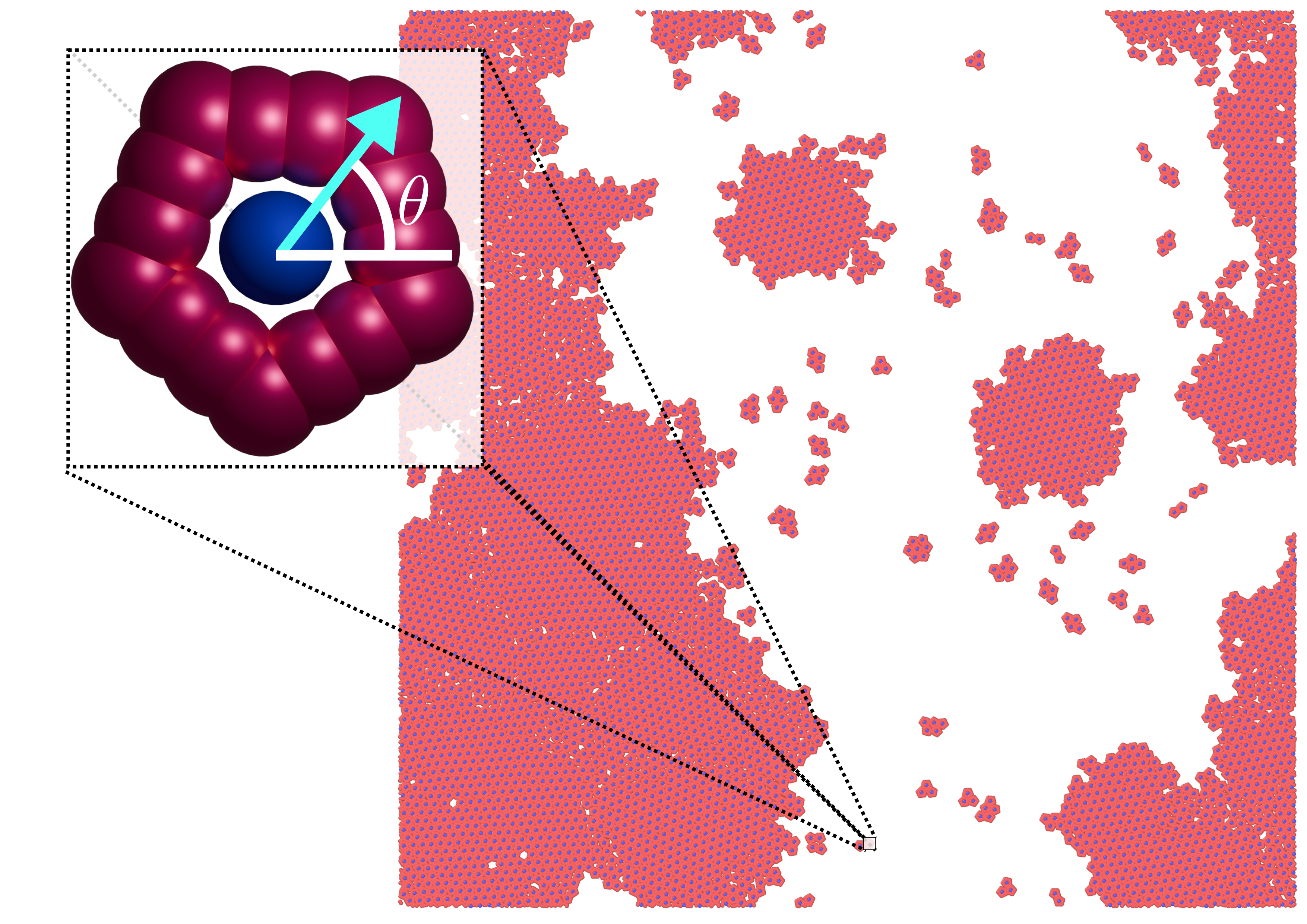}

    \caption{A snapshot of a  suspension of $N = 5000$  pentagons. The zoom shows a single pentagon composed of particles, with a blue central particle surrounded by 15 red particles forming the pentagonal structure; additionally, a cyan arrow indicates the orientation vector along which activity is  applied.
}
    \label{Fig_pentagon_snapshot}
\end{figure}

 A pentagon consists of $16$ particles: $1$ central active particle (blue in Fig.~\ref{Fig_pentagon_snapshot}) at the center of mass and $15$ passive edge-particles (red in Fig.~\ref{Fig_pentagon_snapshot}), each one with a diameter $\sigma$, arranged along the pentagon's edges (four particles per side) to form a rigid pentagonal shell. These edge-particles are evenly spaced and slightly overlapping, resulting in a relatively smooth contour that approximates a regular pentagon 
 (see Fig.\ref{Fig_pentagon_snapshot}). 
 The edge-particle shell is rigidly connected to the central particle, forming an overall pentagon structure with a radius of $R = 2.5\sigma$ \footnote{It is worth mentioning  that a higher number of particles per edge would  improve the geometric fidelity of the structure, thus becoming closer to a smooth pentagon.  However, the computational cost of such simulations would also  dramatically increase. }. Due to the rigid constraint, the whole pentagon effectively behaves as a single rigid body with its activity applied on the central particle (shown in figure \ref{Fig_pentagon_snapshot}).


Each pentagon's center of mass  and its moment of inertia $I$  \textcolor{black}{(see Supplementary Materials)}
evolves according to an overdamped  Langevin dynamics at constant temperature $T$, via the following translational \eqref{Eq_langevin_Trans} and rotational \eqref{Eq_langevin_Rot} equations:
\begin{equation}
M \frac{d^2\vec{r}}{dt^2} = F_a\vec{n} +\vec{F_{\text{int}}} - \gamma_t \dfrac{d\vec{r}}{dt} + \sqrt{2\gamma_t k_B T}\,\vec{\xi}
\label{Eq_langevin_Trans}
\end{equation}
\begin{equation}
    I \frac{d\omega}{dt}=\tau-\gamma_r\omega+\sqrt{2\gamma_r k_B T}\,\eta
\label{Eq_langevin_Rot}
\end{equation}

In Eq.~\ref{Eq_langevin_Trans} and ~\ref{Eq_langevin_Rot}, $M$ is each pentagon's total mass, $\vec{r}$ its center of mass position, and $\omega$ its angular velocity. The constant self-propelling force $F_a$ is aligned with the vector $\vec{n}=(\cos{\theta_{a}},\sin{\theta_{a}})$  (see Fig.\ref{Fig_pentagon_snapshot}), which starts  at the center of mass and points to one vertex of the pentagon, in such a way that it defines an orientation of the pentagon and evolves rigidly with the pentagon via $\dot{\theta}_{Fa} = \omega$.

The parameters $\gamma_t$ and $\gamma_r$ are the translational and rotational friction coefficients, respectively, 
and $k_B$ the Boltzmann constant. The thermal fluctuations are included  via the independent stochastic processes
$\vec{\xi}$ and $\eta$, following a Gaussian white noise distribution with zero mean and unit variance satisfying $\langle \xi_i(t)\xi_j(t') \rangle = \delta_{ij} \delta(t-t')$ and $\langle \eta(t)\eta(t') \rangle = \delta(t-t')$ \footnote{In our numerical simulations, the thermal fluctuations included via the  stochastic noise are drawn from a uniform distribution rescaled with a prefactor $\sqrt{24}$ in order to increase computational efficiency  as described in \protect\cite{dunweg1991}.}(see footnote).

The interaction terms in $\vec{F_{\text{int}}}=-\vec{\nabla} U(\vec{r})$ (Eq.~\ref{Eq_langevin_Trans}) and $\tau$ (Eq.~\ref{Eq_langevin_Rot}) come from the underlying particle-based structure of each rigid pentagon. The particles that constitutes the pentagonal shell interact with their counterparts of other pentagons via the Imperial Pseudo-Hard Sphere (IPHS, Eq.~\ref{Eq_IPHS}) potential~\cite{Potencial_cite_1}:
\begin{equation}
    U(r)=
    \begin{cases}
        50\left(\frac{50}{49}\right)^{49}\epsilon \left[\left(\dfrac{\sigma}{r}\right)^{50}-\left(\dfrac{\sigma}{r}\right)^{49}\right] + \epsilon, & r <\frac{50}{49}\sigma \\
        0, & r \geq \frac{50}{49}\sigma
    \end{cases}
    \label{Eq_IPHS}
\end{equation}

Therefore, the total translational force on the pentagons due to pair-interactions $\vec{F_{\text{int}}}$ at Eq.~\ref{Eq_langevin_Trans} corresponds to the sum of all  forces acting on each particle of the pentagonal shell projected to the center of mass of the pentagon. Similarly, the torque ($\tau$ in Eq.~\ref{Eq_langevin_Rot}) experienced by the pentagon caused by the interactions is obtained by summing the moments of all inter-particle forces about the center of mass.


Our simulations are performed using a house-modified version of LAMMPS \cite{LAMMPS} We keep the number of pentagons constant ($N=5000$) and vary the number density $\rho_n = N / A_{\text{box}}$ by adjusting the size of the simulation box. 
We simulate a two-dimensional bulk suspension 
in a square box with periodic boundary conditions shown in Fig. \ref{Fig_pentagon_snapshot}. 
The box area is defined as  $A_{\text{box}} = L_xL_y = L^2$, where $L_x$ and $L_y$ are the box dimensions. As $\rho_n$ changes, the area fraction  $\rho = A_p \rho_n$ also varies, with $A_p\approx14.9$ the area of a single pentagon. This area is computed by considering a geometric pentagon with circumradius $r_p=2$, corresponding to the positions of the particle centers, and including the particle radius ($\sigma/2=0.5$).

The initial configuration at different area fractions is generated from a dilute suspension (covered area fraction of $\rho=0.01$) of passive ($F_a=0$) pentagons located in a square lattice. Configurations at higher densities are then obtained by performing an NVT simulation, in which the simulation box is gradually compressed, hence increasing the covered area until the desired target densities ($0.1 \leq \rho \leq 0.8$) are reached.

The translational and rotational diffusion coefficient of the rigid body are defined as $D_t= k_B T/\gamma_t$ and $D_r= k_B T/\gamma_r$, respectively, which in our simulations satisfy the Stokes-Einstein relation $D_r=3D_t$. The thermal energy is set to be $k_BT = 1.5$, the friction coefficient $\gamma_t=100$ (being large enough to consider the overdamped regime) and  $\sigma=\epsilon=1$.

We explore area fractions in the range $0.1 \leq \rho \leq 0.8$. For $\rho > 0.8$, the system approaches the maximum packing density, where simulations become computationally expensive due to a dramatic increase of relaxation times \cite{Graeme_2022}.
The integration time step is set to $\Delta t = 0.001$, except for the low area fractions ($\rho=0.1 \text{ and } 0.2$\ro{)}, where a larger time step $\Delta t = 0.005$ is employed, to speed up the system relaxation. Simulations are performed up to a total time of $t_{\text{sim}}\approx47.5\cdot10^{3}$, and for area fractions of $\rho=0.1 \text{ and } 0.2$, $t_{\text{sim}}\approx96.5\cdot10^{3}$ and $t_{\text{sim}}\approx156.5\cdot10^{3}$, respectively.

Besides the area fraction, we  systematically vary  particles' \textcolor{black}{active force}, quantified via the Péclet number  $\text{Pe} = 3\nu \tau_r^{\text{id}} / \sigma$, where $\nu = F_a D_t^\text{id} / k_BT$ is the self-propulsion velocity, $\tau_r^{\text{id}} = 1 / D_r^{\text{id}}$ and $D_t^\text{id}$ the reorientation time and translational diffusion coefficient, respectively, of an ideal pentagon \textcolor{black}{(indicated by the ``id'' label)}, i.e. a non-interacting pentagon defined by removing the terms $\vec{F_{int}}$ and $\tau$ from Eq.~\ref{Eq_langevin_Trans} and ~\ref{Eq_langevin_Rot}, respectively \textcolor{black}{(see \textcolor{black}{Supplementary Materials})}. \textcolor{black}{The Péclet number quantifies the competition between self-propulsion and thermal diffusion, being this a dimensionless control parameter for the activity strength. Here, it is defined using the ideal diffusion coefficients to establish a well-defined control parameter based on single-particle dynamics, unaffected by the collective effects arising from interparticle interactions.} 
We choose to  vary activity  from $\text{Pe}=0$ to $\text{Pe}=120$ \textcolor{black}{(only changing $F_a$)}, and launch a set of simulations at $\text{Pe}=200$ to further test the stability of the system's  behavior at high activities.


\subsection{Analysis tools}


To characterise the structural and dynamical features of these suspensions, several quantities are evaluated. First, the local density, $\phi$, is calculated to determine whether or not dense 
and dilute phases coexist in the suspensions of active pentagons. Following  Refs. \cite{Cates_2014,Jose_2021}, the local density associated to the $i$-th particle is defined as 
    $\phi_{i}=\frac{1}{A_{v,i}}$, 
where $A_{v,i}$ the area of the Voronoi cell of the $i$-th particle (in this case pentagons). This Voronoi tessellation is computed  based  on the positions of the pentagons' center of mass. The local density is rescaled by multiplying it by the area of a pentagon ($A_p$), in such a way that the distribution is constrained between $0$ and $1$, leading to a local area fraction $\tilde{\phi_{i}} = \phi_{i} A_p$. 
The probability distribution function of the local area fraction, $P(\tilde{\phi})$ is defined as:
\begin{equation}\label{Eq_local_density_distribution}
    P(\tilde{\phi}) = \left\langle \frac{1}{N} \sum_{i=1}^{N} \delta(\tilde{\phi} - \tilde{\phi}_i) \right\rangle_t
\end{equation}
where $N$ is the total number of pentagons, $\delta$ is the Dirac delta function, and $\langle\dots\rangle_t$ denotes the time average. In the passive limit ($\text{Pe}=0$), when the system remains structurally homogeneous, $P(\tilde{\phi})$ exhibits a single, well-defined peak located at the system packing fraction. In the active case, this distribution, evaluated once the system has reached a steady state, provides insights on whether the system is homogeneous, or undergoes phase separation, displaying two peaks: one at low packing fraction (indicative of a dilute phase) and another one at higher packing fraction (indicative of a dense phase). Moreover, due to the rescaled $\tilde{\phi}$, the positions of the peaks reflect the actual local packing fraction.

To further characterize the degree of structural order,  the local hexagonal order is calculated
\begin{equation}
\psi_6^k = \frac{1}{n_k} \sum_{j \in N_i} e^{i6\theta_{kj}},
\end{equation}
where $\theta_{kj}$ is the angle between the vector $\vec{r}_{kj}$ (connecting particle $k$ to its neighbor $j$) and the $x$-axis, $N_k$ denotes the corresponding set of nearest neighbors, and $n_k$ is the number of such neighbors.
By construction, the modulus of the hexatic order parameter, $|\psi_6^k|$, approaches unity when its local environment is consistent with an ideal hexagonal lattice, whereas it tends to zero when the local structure deviates significantly from hexagonal symmetry. 
Consequently, the probability distribution of $\psi_6^k$ provides a robust measure of how closely the local structure resembles an hexagonal lattice. 
Averaging over all particles, we compute 
 the instantaneous global hexagonal order parameter,  $\Psi_6(t) = \frac{1}{N} \sum_{k=1}^N \left| \psi_6^k \right|$. To ensure that the system has reached a steady state, the average value $\Psi_6(t)$ is computed as a function of time. The time evolution of $\Psi_6(t)$ serves as a reliable criteria to distinguish whether the system is in steady state, \textcolor{black}{where a plateau is shown}, or undergoing a transient regime \cite{Jose_2022}. 

In addition to  $P(\psi_6)$ and $P(\tilde\phi)$, the arrangement of the pentagons in the suspension  is also characterized by  the radial distribution function (RDF) \cite{Frenkel_1996,hansen_2013},  $g(r)$, which for a two-dimensional system with number density $\rho_n$ is defined as:
\begin{equation}\label{Eq_RFD}
    g(r) = \frac{1}{2\pi r \rho_n} \left\langle \frac{1}{N} \sum_{i=1}^N \sum_{j \neq i}^N \delta(r - r_{ij}) \right\rangle_t,
\end{equation}
where $r_{ij} = |\mathbf{r}_i - \mathbf{r}_j|$ is the distance between the centers of mass of particles $i$ and $j$ and the binning is chosen to be $0.01\sigma$.
The normalisation prefactor $1 / (2\pi r \rho_n)$ corresponds to the probability expected for an ideal gas at the same density $\rho_n$. 


 \begin{figure*}
    \centering
    \includegraphics[width=\linewidth]{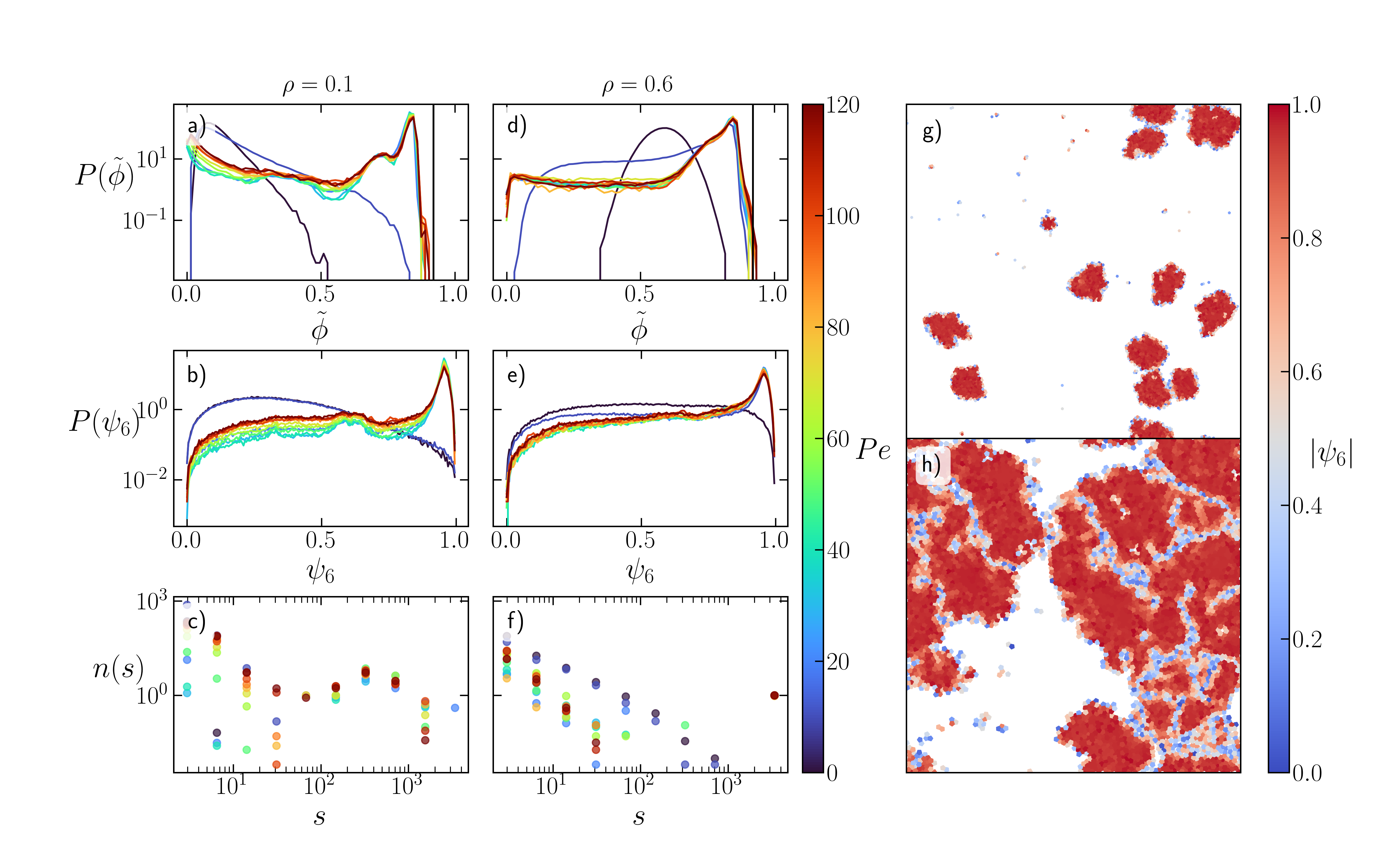}
    \caption{   
    Local density distributions 
    ($P(\tilde{\phi})$) are shown in panels (a) and (d), the bond order parameter distributions ($\psi_6$) are displayed in panels (b) and (e); and cluster size distribution $(n(s))$ is illustrated in panels (c) and (f), for area fractions $\rho=0.1$ (first column) and $\rho=0.6$ (second column). 
In each case, results are presented for different Péclet numbers from  $\text{Pe}=0$ (dark blue) to $\text{Pe}=120$ (dark red). Panels (g) and (h) show snapshots of the suspension\textcolor{black}{, both with Pe=50,} at area fractions $\rho=0.1$ and $\rho=0.6$, respectively; the color of each pentagon indicates the magnitude of $\psi_6^k$.}
    \label{Fig_Local_density_Q6}
\end{figure*}


As   previously reported by \cite{Moran_2022,moran2022particle},  suspensions of 2D active polygons exhibit a cluster phase (micro-phase separation), as can be seen in Fig.~\ref{Fig_pentagon_snapshot}, in addition to MIPS. To quantitatively distinguish these regimes and to gain a deeper insight of the different phases, we evaluate all clusters in the system and compute   
the cluster size distribution, $n(s)$. 
To identify all clusters, we define a distance cutoff within which pentagons can be considered as belonging to the same cluster. Consequently, the smallest considered cluster consists of 
single pentagons. Usually, whenever particles are circular, this cutoff is chosen from the first minimum of the radial distribution function. However, in this case, this distance does not turn out to be too informative, since the pentagons have a non-negligible area. Based on geometrical grounds, we choose $r_\text{c}=5\sigma$.  Given that the effective outer pentagon radius is $R=2.5\sigma$ (distance from the central particle to a vertex particle including its steric radius), setting the cutoff to twice this distance ensures that pentagons in direct contact (vertex–vertex, side–vertex, or side–side interactions) are assigned to the same cluster. Therefore, all the interacting nearest neighbors are included in the same aggregate \textcolor{black}{(see \textcolor{black}{Supplementary Materials})}. 



Self-propulsion  enhances particles' motility, leading to both  anomalous diffusion and dynamical heterogeneity,  i.e. the coexistence of fast-moving and slow-moving particles. 
In order to quantify how pentagons' mobility  changes with  activity and  density, we compute the Mean Square Displacement ($\langle \Delta r^2\rangle$) together with the Non-Gaussian parameter ($\alpha_2$) of the pentagon's center of mass.

On one hand, the Mean Square Displacement~\cite{Jose_2022} characterizes the spatial displacement of particles over time
\begin{equation} \label{Eq_MSD}
    \langle\Delta r(t)^{2}\rangle = \langle |\vec{r}(t) - \vec{r}(0)|^2 \rangle,
\end{equation}
where $\vec{r}(t)$ is the position of the pentagon  center of mass at time $t$ and $\langle\dots\rangle$ denotes an average over all particles.  The MSD follows a scaling relation of the form $\langle \Delta r^2(t) \rangle \sim  t^{\beta}$, where $\beta = 2$ corresponds to the ballistic regime and $\beta = 1$ to normal diffusion. Deviations from these cases indicate anomalous dynamics, which may be sub-diffusive dynamics for $\beta < 1$, typically arising from trapping events or molecular crowding, and super-diffusive dynamics for $ 1<\beta<2$, which is a signature of active transport.

On the other hand, the Non-Gaussian parameter ($\alpha_2$) measures the deviation of  particles' displacement distribution from a purely Gaussian distribution~\cite{Vorselaars_2007,Ginelli_2010,song_2019,Jose_2022}. In a two-dimensional system, this parameter is calculated via:
\begin{equation} \label{Eq_non_gaussian_parameter}
    \alpha_2(t) = \frac{\langle \Delta r^4(t) \rangle}{2 \langle \Delta r^2(t) \rangle^2} - 1,
\end{equation}
where  $\Delta r(t)$ denotes the displacement of a particle over a time interval $t$, i.e., $\Delta r(t) = |\vec{r}(t) - \vec{r}(0)|$ . While $\alpha_2=0$ corresponds to an ideal Brownian motion resulting in a Gaussian distribution of displacements, larger values ($\alpha_2>0$) imply dynamical heterogeneity, as observed during MIPS in suspensions of repulsive spherical ABP, where  particles' displacements differ from the average \textcolor{black}{producing a bimodal distribution}: particles are slow in the dense phase and fast in the gas phase.

\section{Results}\label{sec: results}


To start with, we study the effect of activity and density on the system's phase behavior. 
At first glance, as shown in the snapshot of Fig.~\ref{Fig_pentagon_snapshot}, dense phases may be formed. Therefore,  to identify the system's phase, the local density distribution ($P(\tilde{\phi})$), together with the $\psi_6$ distribution, is computed. 
In Fig.~\ref{Fig_Local_density_Q6} we report the local density distribution  ($P(\tilde{\phi})$) for densities $\rho=0.1$ \ (panel a) and $\rho=0.6$ (panel d), each having various cases with $\text{Pe}\in[0,120]$ (as indicated by the color code). 
Both panels clearly show the different phases  present in the system, ranging from a uniform phase (with a broad peak around the corresponding $\rho$ of the suspension) to coexisting phases for $\text{Pe}>10$ (with two peaks at high/low densities). \textcolor{black}{See \textcolor{black}{Supplementary Materials} for a complete overview of the local density distribution and $\psi_6$ for all studied densities and activities}. \textcolor{black}{Also, in the passive case ($\text{Pe}=0$, dark blue in Fig.~\ref{Fig_Local_density_Q6}a and d), while both local density distributions display a well-defined peak around the average packing fraction of the system, their decay behavior is significantly different. At $\rho=0.6$ (dense fluid regime), geometric frustration and interlocking effects restrict pentagon mobility, leading to short-lived clusters and symmetric local density fluctuations (appearing almost quadratic on a logarithmic scale). Conversely, at $\rho=0.1$, pentagons move more freely in a low-density fluid regime where transient encounters form short-lived aggregates, causing the existence of higher density regions (up to $\tilde{\phi}\approx 0.5$) and the exponential-like decay at $P(\tilde{\phi})$ reflecting the statistical probability of finding denser regions (see Supplementary Materials).}

We have also computed the  distribution of  the local hexagonal order, $ \psi_6$, as shown in Fig.~\ref{Fig_Local_density_Q6} panels b and e.
Apart from the case without a dense phase, the maximum value of this parameter is  close to 1, indicating that the internal structure of the dense phase mostly contains pentagons with six neighbors.
In order to clarify these results, two snapshots are included in panels g and h, where the color code of each pentagon corresponds to its $ \psi_6$ value (from zero, blue, to one, red). Clearly,  either in a 
cluster phase (low density, panel g) and in the MIPS (higher density, panel h), 
 bulk pentagons exhibit a high $\psi_6$ value. The internal structure of  clusters (panel g) is more ordered in terms of the hexagonal order parameter, since there are no grain boundaries within them. 
This contrasts with the case observed at a density of $\rho=0.6$ (panel h), which exhibits regions with low $\psi_6$, corresponding to structural defects  giving rise to grain boundaries. 
Nevertheless, the motility induced phase separation observed in this system is not fully equivalent to the MIPS typically reported in suspensions of active disks~\cite{cates2015motility}. In the present case, the population of particles in the dilute region is significantly reduced, due to the geometrical difficulty for pentagonal particles to escape from the dense region into the dilute one. Consequently,  the system more closely resembles an arrested MIPS rather than a conventional MIPS, since it is even possible to observe a single large cluster containing all  particles of the system.


To clearly distinguish the cluster phase from the MIPS one (Fig.~\ref{Fig_Local_density_Q6} panels g and h, respectively), we now compute 
the cluster size distribution ($n(s)$)
, shown in Fig.~\ref{Fig_Local_density_Q6} for densities $\rho=0.1$ (panel c) and $\rho=0.6$ (panel f) using the same color code for $\text{Pe}$ as the previously presented quantities. 
As expected, at high density $\rho\in\left[0.6, 0.8\right]$ the $n(s)$ with $\text{Pe}\neq0$ 
displays two peaks 
\textcolor{black}{(see \textcolor{black}{Supplementary Materials} for a whole view of the cluster size distribution at several densities)}. The first, larger, peak occurs at $s=1$ (\textcolor{black}{single pentagons}) and decays with increasing $s$, \textcolor{black}{being pentagons pairs the} 
small\textcolor{black}{est} clusters formed due to short-time collisions. The second, sharper peak at $s\approx 3000$ represents a unique large aggregate 
(see Fig.~\ref{Fig_Local_density_Q6} panel f). This reflects the coexistence of transitory clusters, 
moving in the dilute phase and a large cluster representing the dense phase, as in MIPS. 
At $\text{Pe}\geq30$, the number of small clusters is reduced and the dense region becomes larger: this highlights the role of activity in driving the MIPS formation.

When $n(s)$ is computed at low density 
(see Fig.~\ref{Fig_Local_density_Q6} panel c), in the range $0.1 \leq \rho \leq 0.4$,  two peaks are displayed for all  explored activities $\text{Pe}>20$. In contrast to the high density cases, both maxima are broad, implying a wider distribution of cluster sizes.
The first peak is at $s=1$ and decays as $s$ increases, corresponding to the analogous found at high density. The second peak shown at $s>200$, is shifted to smaller $s$ compared to the high density case, reveals the presence of several-sizes clusters.
While for $ 0.1 \le \rho \le 0.3$ the cluster phase is clearly identified, as $\rho$ increases the distribution becomes gradually noisier \textcolor{black}{(see \textcolor{black}{Supplementary Materials})}, particularly at densities $\rho=0.4$ and $\rho=0.5$. This behavior indicates the progressive transition from the cluster phase to a MIPS at higher $\rho$.

Nevertheless, it is striking that small clusters are present at all values of density and activities, due to pentagons' anisotropy and their corrugated shape.
When two pentagons interact and the activity is fixed at a vertex,  kinetic binding is much stronger thar disk particles since it is more difficult for them to rotate and, thus, change their direction of motion  to escape from one another.
This is because the rotation of one will cause the other to rotate, forming pairs or groups of three pentagons  rotating around their own axes without coming apart for a certain period of time \textcolor{black}{(see video at \textcolor{black}{Supplementary Materials})}. The kinetic trapping is enhanced due to the corrugated sides of pentagons, allowing a hard interlocking between them. 


\begin{figure}[h!]
    \centering
    \includegraphics[width=1\linewidth]{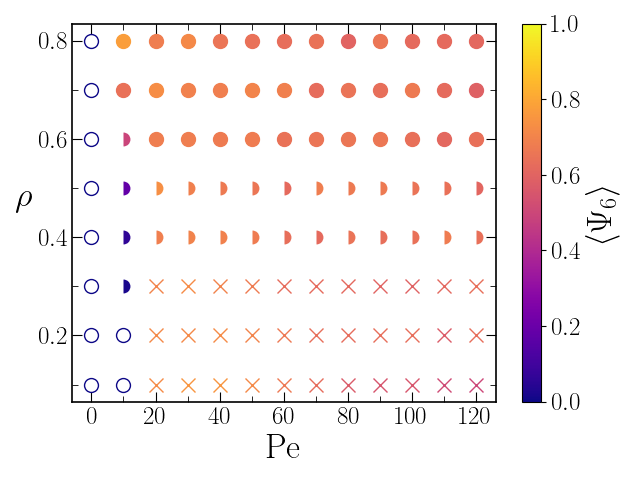}

    \caption{Phase diagram of a two-dimensional suspension of $N=5000$ active pentagons at different $\text{Pe}$ number and density. Empty dots: homogeneous phase; cross: cluster phase; filled dots: MIPS; half dots: not clear difference between cluster phase and MIPS. Moreover, the color used encodes the average over particles and time of the $\psi_6$.
}
    \label{Fig_Phase_Diagram}
\end{figure}

Having all tools in place, we now perform a set of simulations at several densities and activities and report the steady-state results in the state diagram reported in Fig.~\ref{Fig_Phase_Diagram}, where empty dots represent the homogeneous phase, cross-like symbols the cluster phase (resembling the micro-phase separation observed in  \cite{moran2022particle} for several polygons) and the filled dots the MIPS phase. 
\textcolor{black}{Half-filled symbols have been used whenever a clear distinction between the cluster phase and MIPS is ambiguous. This classification is based on the fraction of particles in the largest cluster alongside the peak width normalized by the weight-average cluster size (see Supplementary Material).
} The color code  in the figure encodes the $\langle \Psi_6 \rangle_t$ value (averaging over all pentagons and time).

The state diagram shows that even at low activity ($\text{Pe}\approx20$),  pentagons form dense structures like dense clusters ($\rho<0.4$)  or MIPS (with $\rho>0.5$). \textcolor{black}{While at intermediate regimes, as $\rho=0.4 \text{ or } 0.5$ the difference between MIPS and cluster phase is not clear (see Supplementary Material).}
\begin{figure*}
    \centering
    \includegraphics[width=1.0\linewidth]{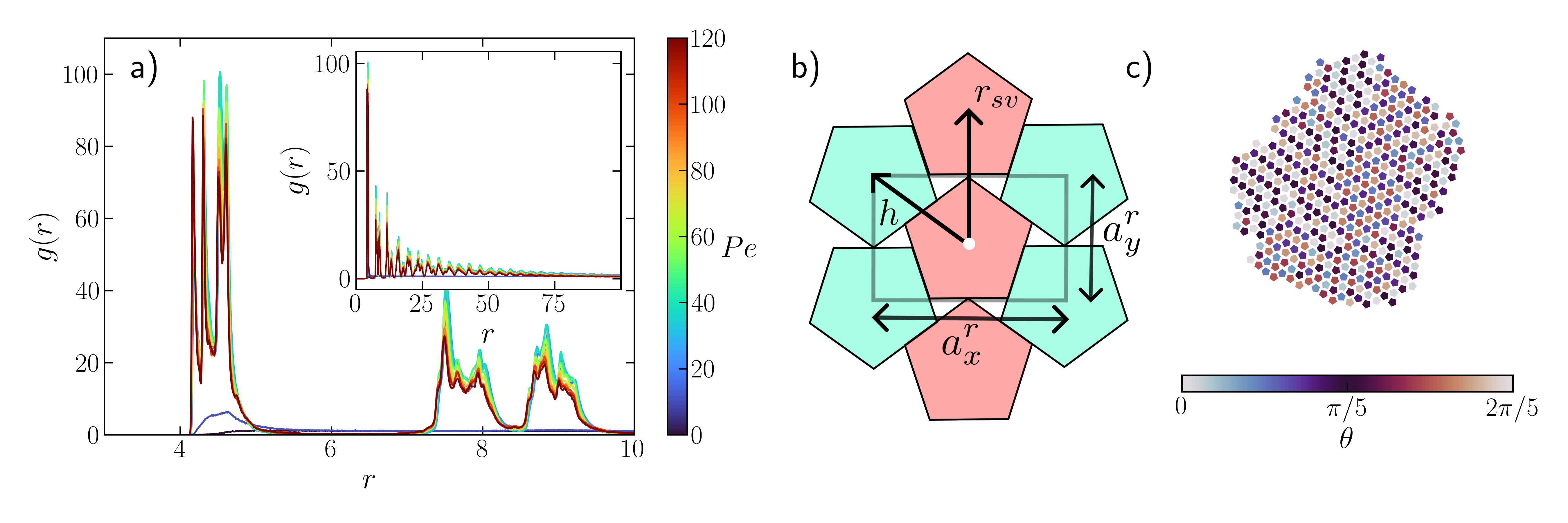}

    \caption{Panel a: Radial  distribution function ($g(r)$) varying the $\text{Pe}$ (from 0, dark blue, to 120, dark red) for a suspension which area fraction is  $\rho=0.1$ is shown. The first-neighbor peak and the next two coordination shells are highlighted. The inset displays the same $g(r)$ over longer distances for the same $\text{Pe}$ values, revealing the long term structure. Panel b: illustrates the densest packing arrangement that a pentagon suspension can form \cite{henley1986sphere,kuperberg1990double,hales2016packings} an ice-ray structure. Panel c: snapshot of a cluster from the same simulation of panel a with $\text{Pe}=40$. The color of each pentagon represents the angle between the positive x-axis and the pentagon vertex closest to that axis (not necessarily the active force angle), indicating crystalline order. Since a regular pentagon presents five-folded symmetry the colorbar used is cyclic ($\theta=0$ equivalent to $\theta=2\pi/5$)  
    }
    \label{Fig_RDF}
\end{figure*}
This singular behavior is caused by two competing factors: the anisotropic shape of the pentagons and their persistent motion due to their constant active force.  We observe that all detected phases are stable against activity: when simulating the system at all densities and a very high activity ($\text{Pe}=200$) the phases are not affected.

Unlike  spherically shaped active particles, when particles are pentagon-like, the effect of activity seems much less pronounced in establishing the phase behavior (at least within the studied activity range)~\cite{Cacciuto_2016,  cates2015motility}.  Therefore, the phase (whether cluster or MIPS) seems to depend more on density than on activity, given that a weak activity is enough to promote a compact structure. 
 




Given the nature of pentagons— the simplest regular polygon not able of tiling the two-dimensional plane~\cite{henley1986sphere,kuperberg1990double,hales2016packings}— we analyse the  structure of the dense phases at high $\rho$, whether cluster or MIPS, examining  the radial function, $g(r)$, together with its crystalline order.


 Figure~\ref{Fig_RDF} panel a represents  the radial  distribution function at $\rho=0.1$ and all  studied $\text{Pe}$  (\textcolor{black}{a more extensive plot is displayed in the Supporting Material}).  
When $\text{Pe}>10$, we observe a long range structure (in the inset) with a high  first  peak, meaning that the inner structure of the clusters tends to be crystal-like rather than gas-like.  
On the other side, when $\text{Pe}\leq10$ the behavior resembles that of a liquid-like, showing a wide first peak at short distances. 


Focusing on the first-neighbors peak (main panel in Fig. \ref{Fig_RDF}a), we can better characterise the dense-phase structure.  The first peak  splits in four peaks at several distances: this indicates  that there are four dominant distances of a pentagon's  neighbors  at distances $r\approx4.17,\, 4.31,\, 4.53$  and $4.61$.

The arrangement of the neighbors could be estimated either  using the geometric radius of the pentagon ($r_p=2$)  or via the geometric radius together with the area of the particles outside the pentagon ($\sigma/2=0.5$). It is to be expected that the minimum distance between pentagons is the distance at which they share sides, i.e. side-to-side $r_{ss}$, which is twice the pentagon apothem ($a_p$). However, neither using the geometric radius nor the radius in combination with an additional particle radius gives a peak  matching the one measured in the radial distribution function, being those distances approximately $3.23$ and $4.04$. 
Both underestimate the distance due to the microscopic shape of the pentagons. When the suspension's packing fraction is high, the pentagons' microscopic details  become important.
Since the surface  is not totally flat, but corrugated,  $r_{ss}$ is  not just the distance between side-by-side pentagons, 
but rather it should take into account when  particles of one pentagon fit into the gaps between  particles of the neighboring one. Therefore, the following correction $\Delta r_{\text{gap}}$ shall be introduced:
\begin{equation}
    \Delta r_{\text{gap}}=2a_p+\sqrt{\sigma^2-d^2/4}
\end{equation}
with $d$ being the distance between particles on the edge.
At the same time, interlocking of pentagons, like gears, causes a displacement in the direction parallel to their sides, of magnitude $d/2$, renaming $r_{ss}$ as $r_{ss}^{d/2}$. Consequently, the real distance between  pentagons in  side-to-side position is 
\begin{equation}
    r_{ss}^{d/2}= \sqrt{4a^2+\sigma^2+4a\sqrt{\sigma^2-d^2/4}}\approx4.17,
\end{equation}
which matches perfectly the distance observed at the first peak of the radial  distribution function. Thus, when the packing fraction of the cluster/MIPS is high, the microscopic structure of the pentagon becomes important. 

Having identified the first of the four degenerate peaks of $g(r)$ with the distance given as $r_{ss}^{d/2}$, it is expected   the remaining peaks to coincide with the distances corresponding to side-to-vertex ($r_{sv}$) or vertex-vertex ($r_{vv}$). 
By considering $r_{sv}$ and taking into account the corrugated surface,
peaks at $r\approx4.53$ and $\approx4.61$ are identified \textcolor{black}{(See \textcolor{black}{Supplementary Materials})}. The former, $r\approx 4.61$, is the furthest maxima of the first peak observed in  the $g(r)$, so that the vertex-to-vertex configuration of pentagons is not represented in the radial  distribution function, being not too probable to detect it. 
The remaining unidentified peak at $4.31$ between $r_{ss}^{d/2}$ and $r_{sv}^{c}$ coincides with the $r_{ss}^{d/2}$ value, shifting the fit by a distance of $3d/2$ rather than just $d/2$, as observed for the first peak \textcolor{black}{(See \textcolor{black}{Supplementary Materials})}.

The radial distribution function and its multiple maxima at first neighbors highlight the inner structure of the pentagon's crystal-like phase.

Figure~\ref{Fig_RDF} panel b reports a schematic picture of an ice-ray structure: this is  the densest packing of pentagons in a two-dimensional Euclidean space \cite{henley1986sphere,kuperberg1990double,hales2016packings}, and consists of  a rectangular lattice with lattice parameters \cite{schilling2005monte}
\begin{equation}\label{Eq_lattice_vectors}
    a_y^r=r_p\frac{5+\sqrt{5}}{4}, \hspace{1cm} a_x^r=\frac{3}{2}r_p\sqrt{\frac{5+\sqrt{5}}{2}}.
\end{equation}

\noindent It corresponds to an arrangement of pentagons consisting of parallel columns of inverted pentagons (see Figure~\ref{Fig_RDF} panel b. 
Pentagons on the same column are separated a distance that corresponds to a vertex-to-side configuration, while pentagons between columns are shifted side-to-side.

Figure~\ref{Fig_RDF} panel c reports 
 a close-up of a cluster formed at a density of $\rho=0.1$ and $\text{Pe}=40$, with the color-code representing the angle $\theta$ formed by the pentagons relative to the $x$ axis taking into account the five-fold symmetry and considering $\theta=0$ \textcolor{black}{when the} vertex of the pentagons \textcolor{black}{is parallel to the x-axis}. Clearly, we observe the  ice-ray structure with  defects. 
  The pair of colors  (dark-light or blue-brown) reveals the parallel columns of inverted pentagons, i.e. rotated by $\pi/5$, together with different domains separated by grain boundaries or dislocations. 

The ice-ray-like structure is represented also in the $g(r)$, which can be quantified using the lattice vectors of Eq.~\ref{Eq_lattice_vectors}. 
The distance between pentagons of the same column (Fig.~\ref{Fig_RDF} b red-red) is the one given by $r_{sv}$, while the distance with neighbors of different columns (Fig.~\ref{Fig_RDF} b red-turquoise) may be calculated by the equation
\begin{equation}
    h=\frac{r}{4}\sqrt{\frac{77+9\sqrt{5}}{2}},
\end{equation}
which, using $r=r_p+\sigma/2$, leads to a distance of $h\approx4.35$. This is close to the one measured in the radial distribution function  corresponding to a shifted side-side distance of $4.31$. Thus, the radial  distribution function highlights the inner structure of the densest phases on the suspension, providing a fingerprint of the spontaneously emerging ice-ray-like structure.

This structure is also present on the second coordination disk (see \textcolor{black}{Supplementary Materials}) of $g(r)$ (included in Fig.~\ref{Fig_RDF} a),
confirming the underlying ice-ray structure and, consequently, the multi-peak splitting in the $g(r)$ for the next-nearest neighbors is also direct consequence of the anisotropy of the ice-ray structure.




Depending on the different collective behaviors,  whether clusters or MIPS-like, pentagons display different dynamics. Unlike disks,  pentagons cannot freely rotate: 
due to their anisotropic shape, once   trapped in a dense phase, the probability of leaving it again is quite low. 
However, since their activity is always on, pentagons induce net forces on the aggregates (estimated by adding all active forces and torques within the clusters) causing their motion. 


  \begin{figure}[h!]
    \centering
    \includegraphics[width=1\linewidth]{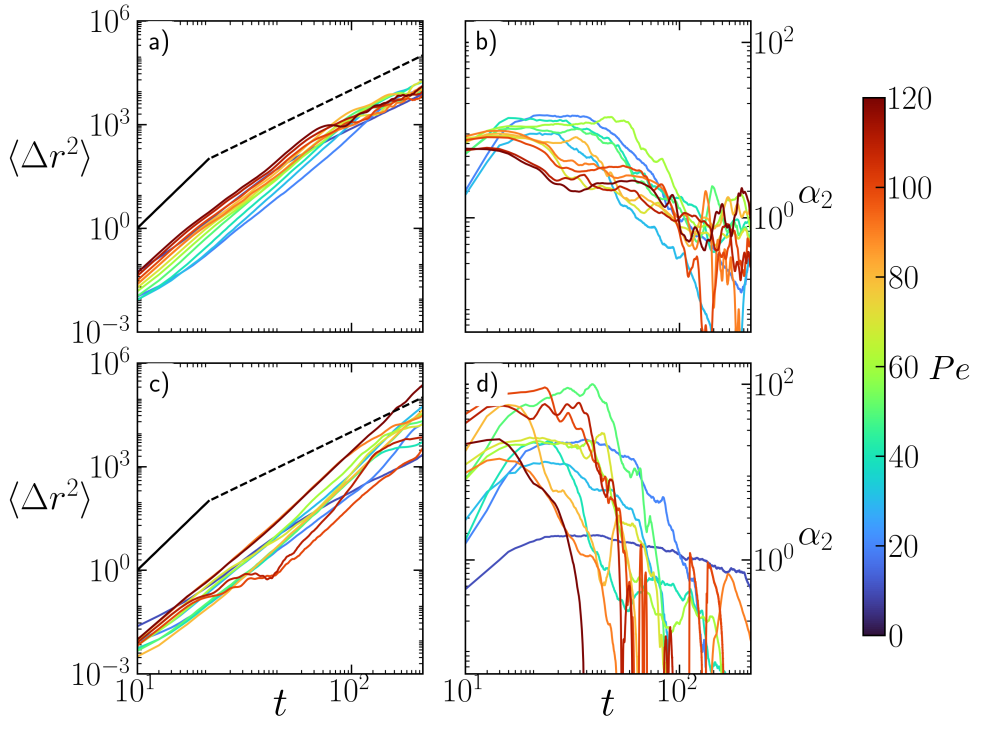}

    \caption{Mean Square Displacement $\langle \Delta r^2\rangle$ (left column, panels a and c) together with the Non-Gaussian parameter (right column, panels b and d) for the packing fractions $\rho=0.1$ (upper row, panels a and b) and $0.6$ (lower row, panels c and d) varying the $\text{Pe}$ number in the range $0$ (dark blue) to $120$ (dark red). In panels a and c, where the MSD is shown two guide lines are included: $\Delta r^2\sim t$ (dashed line) and $\Delta r^2\sim t^2$ (black solid line).
}
    \label{Fig_MSD_NGP}
\end{figure}

Figure \ref{Fig_MSD_NGP} represents the Mean Square Displacement ($\langle\Delta r^2\rangle$) and Non-Gaussian Parameter ($\alpha_2$) as a function of time at two different area fractions   ($\rho=0.1$, upper row, and $0.6$, lower row) in a broad activity range  (from $\text{Pe}=0$ to $120$, as reported in the color code). 
These dynamical properties give us information on the system's diffusion and the presence of dynamic heterogeneities. 



The Mean Square Displacement computed by averaging over all pentagons, is shown in Fig.~\ref{Fig_MSD_NGP} panels a ($\rho=0.1$) and c ($\rho=0.6$). 
Two regimes can be distinguished: a ballistic one ($\langle\Delta r^2\rangle\sim t^{2}$, solid black line as a guide) and a diffusive one ($\langle\Delta r^2\rangle\sim t$, black dashed line as a guide).

 In the passive limit (dark blue in Fig.~\ref{Fig_MSD_NGP}) at both densities $\langle\Delta r^2\rangle$ shows a short ballistic regime and rapidly relaxes to a diffusive one.
 Conversely, when  activity is  low,  the ballistic regime lasts for longer than in the passive case. 
 While at low density it lasts up to $t\approx 500$, at high density it seems to be even longer than the simulated time
 (this is not observed in an ideal suspension of active pentagons \textcolor{black}{(see \textcolor{black}{Supplementary Materials})}.

In all active cases, the behavior is reminiscent of the cluster/MIPS motion, providing an initial insight into the dynamics of these dense phases, which display translation and rotation.
At low density, where the cluster phase appears, the MSD mostly resembles that of the clusters
since  pentagons are enclosed within them and cannot move freely beyond the cluster’s own motion. 
Hence, we conclude that 
clusters exhibit a long-time diffusion, with a long ballistic regime. 

However, in the diffusive regime, the MSD is not entirely linear in Fig.~\ref{Fig_MSD_NGP} a, but exhibits periodic fluctuations, typical outcome of rotating particles, \textcolor{black}{being this another evidence of the activity induced rotation and translation that clusters often exhibit}.
The arrangement of pentagons within a cluster causes its translation due to an imbalance between the preferred directions of the pentagons, resulting in a non-zero average; the same applies to the total torque acting on a cluster.
Therefore,  we can safely assume that the larger the cluster, the closer the average of the active forces will be to zero. Moreover, since the mass increases, the dynamics will also slow down. This is supported by the results obtained in the MIPS simulations, where the dense phase is much larger than that observed in the \textcolor{black}{cluster phase} and its movement and rotation is slower.

As suggested by the $\alpha_2$  in Figure ~\ref{Fig_MSD_NGP} b and d, while in the passive limit ($\text{Pe}=0$), the suspension remains homogeneous, showing close-to-zero values, a low activity is sufficient to drive the suspension, at high density, into aggregated states  (such as clusters or MIPS).
This transition is reflected in the particle dynamics, since $\alpha_2$ takes values significantly different from zero, indicating the appearance of heterogeneous mobility between the fast-moving dilute phase and the slow particles subject to steric interactions within the clusters.

Moreover, in low density systems at long times (Fig.~\ref{Fig_MSD_NGP} b), we observe that $\alpha_2$ does not fully relax to zero. 
This incomplete decay shows a persistent dynamic heterogeneity, suggesting that  particles remain kinetically trapped within the dense domains over the simulated timescales. Nevertheless, in high density systems at long times (Fig.~\ref{Fig_MSD_NGP} d)  $\alpha_2$ decays to close-to-zero values when $\text{Pe}>10$: this might indicate that all  pentagons have been integrated into the clusters, being the movement homogeneous again with only a slow motion being present overall.

\section{Conclusions} \label{sec: conclusions}

We study the collective behavior of a two-dimensional suspension of active repulsive pentagons (interacting via an IPHS potential) while  varying the $\text{Pe}$ (via the active force) and the packing density.
Both structural and dynamical properties of the system have been analyzed to construct a phase diagram as a function of density and activity.

In our simulations, each active pentagon (characterized by slightly corrugated edges and round vertex) is composed by $16$ particles: a central particle, at the center of mass, surrounded by $15$ slightly overlapping particles forming a rigid outer shell (rigidly bonded together and to the central particle).
Each pentagon equations of motions follows 
a Langevin dynamics in the Brownian limit. 

We observe that 
 the active pentagons' anisotropic form, enhanced by the non-smooth edges of the pentagons, drives the suspension towards a phase coexistence of dense and dilute regions. \ro{T}his phase separation happens \textcolor{black}{at lower activities than the ones reported in a two-dimensional suspension of ABP~\cite{Cates_2014}}.
In the coexistence regime, two different states emerge: a cluster phase and a motility induced-phase separation (MIPS).
The criterion to further distinguish between these states is based on the morphology of the dense regions: the cluster phase consists of  finite-sized aggregates, whereas MIPS is characterized by a single macroscopic dense area, i.e. a single large cluster.

While MIPS is only observed at  high density and activity, the cluster phase is found even under conditions of very low density and activity. This highlights that once  activity is even slightly on, the phase behavior of the suspension is not dependent on the strength of the active force, but rather on the density.

The tendency of the suspension to build clusters arises from the interplay between the pentagonal geometry of the rigid bodies and the persistent  self-propelling force, which is strictly aligned with a single vertex.
To change the self-propelling force direction, a pentagon shall rotate completely, and in crowded environments its rotational diffusion is severely reduced. This leads to a kinetic arrest of the pentagons which, coupled with the corrugated edges,  promotes the interlock, thus stabilizing the clusters.

The inner structure of the dense regions is also characterized via radial distribution function alongside with the hexagonal order parameter, based on Voronoi neighbors.
The high values of this orientational order parameter confirm that  pentagons spontaneously self-assemble into a highly ordered hexagonal lattice within the clusters. The local structure formed is an ice-ray pattern, which is the densest two-dimensional packing for pentagons, as predicted by \cite{moran2022particle} for pentagon-like active polygons. 
This spontaneous ordering highlights the critical role played by the microscopic structure of the individual pentagons in the assembly process.

Furthermore, dense phases, whether   finite clusters or  MIPS-like, are not static over time but rather exhibit dynamical behavior.

To further characterize  the dynamics, the Mean Square Displacement and the Non-Gaussian parameter are computed. The pentagons' motion within the suspension shows long-time heterogeneities due to the coexistence of kinetically trapped pentagons in the dense regions (which are slow) together with  moving pentagons in the dilute regions (which are highly active).

The MSD  highlights the collective persistence displayed by the dense aggregates. 
An extended ballistic regime, which  exceeds a single pentagon reorientation time, is observed alongside the emergence of a chiral cluster's motion. 
Both, the translational and rotational movement of the dense regions, are caused by an imbalance of self-propelling forces of the constituent active pentagons leading to non-zero net force and torque.

Our numerical work on the emergent collective behavior in a two-dimensional suspension of active pentagons aims at giving guidance to experiments on such systems. 
Active pentagon-like particles can be synthetically created  using various techniques, ranging from a microscopic approach (either chemical~\cite{Palacci_2010,Ginot_2015} or magnetic~\cite{Calero_2020}) to a macroscopical one, potentially based on 3D printed granular particles~\cite{farhadi2018dynamics,xiao2022probing,Lin_2026} excited by air-flows~\cite{Ojha2004,Abate_2006,Daniels_2009}.
In all these cases, activity can be induced by breaking the particles' symmetry, in terms of its composition or structure. 
Work is in progress to experimentally unravel the features of a collection of pentagons, either passive or active, whose effective interactions resemble an effective short-range attraction.

\section{Supplementary Material}\label{sec:suppmat}

The Supplementary Material reports the moment of inertia for an ideal suspension of pentagons; a thorough study of the cluster size distribution; a  complete overview on local density, bond order parameter, cluster size distribution and radial distribution function;  besides few illustrative videos. 

\section*{Acknowledgments}
C.V. acknowledges funding from IHRC22/00002, PID2025-170781NB-C21 and Proyecto PID2022-140407NB-C21  by MCIN/AEI /10.13039/501100011033 and FEDER, UE. M.R.-G. and R.F.-Q.G. acknowledge support from Ramón y Cajal program (RYC2021-032055-I) funded by MCIN/AEI/10.13039/501100011033 and by European Union NextGenerationEU/PRTR, a Research Grant from HFSP (Ref.-No: RGEC33/2024) with the award DOI ({10.52044/HFSP.RGEC332024.pc.gr.194170})
and grant PID2023-147067NB-I00 funded by MCIU/AEI/10.13039/501100011033 and by ERDF/EU. Google Gemini was used for checking the spelling and grammar of the manuscript.
H.X.,  and W.L. would like to acknowledge the funding from the National Science Foundation grant CMMI-2519512.


\bibliography{References_correct}

\end{document}